\documentclass[11pt,twoside]{article}
\usepackage{graphicx}
\usepackage{amsmath}
\usepackage{amssymb}
\usepackage{cite}

\begin{document}
\newcommand{\pst}{\hspace*{1.5em}}

\newcommand{\rigmark}{\em Journal of Russian Laser Research}

\newcommand{\be}{\begin{equation}}
\newcommand{\ee}{\end{equation}}
\newcommand{\bm}{\boldmath}
\newcommand{\ds}{\displaystyle}
\newcommand{\bea}{\begin{eqnarray}}
\newcommand{\eea}{\end{eqnarray}}
\newcommand{\ba}{\begin{array}}
\newcommand{\ea}{\end{array}}
\newcommand{\arcsinh}{\mathop{\rm arcsinh}\nolimits}
\newcommand{\arctanh}{\mathop{\rm arctanh}\nolimits}
\newcommand{\bc}{\begin{center}}
\newcommand{\ec}{\end{center}}

\thispagestyle{plain}

\label{sh}


\begin{center} {\Large \bf
\begin{tabular}{c}
DYNAMICS OF BEC MIXTURES LOADED
\\[-1mm]
INTO THE OPTICAL LATTICE  
IN THE PRESENCE 
\\[-1mm]
OF LINEAR INTER-COMPONENT COUPLING
\end{tabular}
 } \end{center}

\bigskip

\bigskip

\begin{center} {\bf
 M.Yu.~Uleysky$^{1}$, D.V.~Makarov$^{1*}$
}\end{center}

\medskip

\begin{center}
{\it
$^1$ 
V.I.~Il'ichev Pacific Oceanological Institute of FEB RAS
\\ Vladivostok, Russia
}
\smallskip

$^*$Corresponding author e-mail:~~~makarov~@~poi.dvo.ru\\
\end{center}

\begin{abstract}\noindent
We consider dynamics of a two-component 
Bose-Einstein condensate where the components
correspond to different hyperfine states of the same sort of atoms.
External microwave radiation leads to resonant transitions
between the states.
The condensate is loaded into the optical lattice.
We invoke the tight-binding approximation and
examine the interplay of spatial and internal dynamics
of the mixture.
It is shown that internal dynamics qualitatively depends on 
the intra-component interaction strength and
the phase configuration of the initial state.
We focus attention on two intriguing phenomena occurring for certain parameter values.
The first phenomenon is the spontaneous synchronization of Rabi oscillations running
inside neighbouring lattice sites.
Another one is the demixing of the condensate with formation of immiscible 
solitons when nonlinearity becomes sufficiently strong.
Demixing is preceded by the transient regime with highly
irregular behavior of the mixture.

\end{abstract}

\medskip


\section{Introduction}\label{Intro}
\pst

Two-component Bose-Einstein condensates
with linear inter-component coupling 
have become an object 
of considerable interest associated with their non-trivial
dynamical properties, such as onset of self-trapping in Rabi oscillations, $\pi$-oscillations \cite{Raghavan},
bifurcations in Rabi oscillations \cite{Oberthaler1},
miscibility-immiscibility transitions 
induced by Rabi coupling \cite{Malomed,Oberthaler2}, to name a few.
Owing to the close analogy with Josephson transitions, these 
two-component mixtures are often referred to as the Bose-Josephson junctions \cite{Raghavan,Gati}.
They are also considered as advantageous tools for
storage of quantum information \cite{Byrnes}, 
or for manufacturing of high-precision atomic clocks \cite{Clocks}.

Usually, two ways of composing a Bose-Josephson junction 
are considered. The first one is based on the creation of the double-well optical potential.
Then different components relate to the condensate fractions trapped inside different wells,
and the linear inter-component coupling is provided by inter-well tunneling.
The second way is the usage of condensates occupying the same trap
but belonging to different hyperfine energy levels. The 
resonant microwave radiation causes transitions between the levels, providing
linear inter-component coupling.
In both these situations atomic ensembles 
are tightly confined inside the optical trap, and
spatial expansion of the condensate is limited.
Elimination of the BEC expansion allows one to describe the inter-component
transitions by means of the classical equations of motion corresponding
to a pendulum-like Hamiltonian \cite{Raghavan,Oberthaler1}.
Meanwhile, it is reasonable to examine the case when
the mixture of two linearly-coupled hyperfine states
is loaded into the optical lattice and
allowed to expand along many potential wells, that is, spatial BEC expansion cannot be eliminated.
This system can be regarded as an array of Bose-Josephson junctions \cite{Science}
and considered as a promising way for creating large-scale entangled states \cite{Calarco,Bloch}.
However, it is well known that 
interplay between external (i.e. spatial) and internal (corresponding to level transitions) degrees of freedom can essentially 
complexify atom dynamics even in the absence of interactions between atoms.
Earlier studies in the semiclassical regime for
the spatial atomic dynamics revealed many intriguing phenomena like
chaotic Rabi oscillations \cite{PRA01}, chaotic atomic
transport and dynamical fractals \cite{JETP03,PLA03,PU06,PRA07}, 
occurrence of L\'evy flights and anomalous diffusion
\cite{JETPL02,PRE02,JRLR06},
synchronization between spatial and internal dynamics
\cite{PRA05}.
A detailed theory of chaotic atomic transport of
atoms in a laser standing wave has been developed in Refs.~\cite{PRA07,PRA08,EPL08}.
Obviously, presence of interactions between atoms should cause additional complications
of the underlying physics.

In the present work we consider the BEC mixture loaded into the lattice in the presence of interactions between atoms. 
Our main goal is to describe the main regimes
of spatial and internal dynamics of the BEC mixture, and to study how dynamics of Rabi oscillations changes
with increasing the interaction strength.
The paper is organized as follows.
In the next section we represent the model considered 
and give brief description of the dynamics when tunneling between neighbouring lattice sites is negligible and there 
is no spatial condensate expansion.
Section \ref{Spatial} describes spatial BEC dynamics when tunneling becomes significant.
In section \ref{Internal} we examine how spatial dynamics influences inter-component transitions.
The results obtained are discussed in Summary.

\section{General theory}\label{General}

\subsection{Basic equations}\label{Basic}
\pst

Within the mean-field approximation,
the condensate dynamics is governed by the coupled Gross-Pitaevskii equations
\begin{equation}
 \begin{aligned}
  i\hbar\frac{\partial\Psi_1}{\partial t}&=
  -\frac{\hbar^2}{2m}\frac{\partial^2\Psi_1}{\partial x^2}
  +U(x)\Psi_1 + g_1|\Psi_1|^2\Psi_1 + g_{12}|\Psi_2|^2\Psi_1
  -\frac{\hbar\Omega}{2}\Psi_2,\\
  i\hbar\frac{\partial\Psi_2}{\partial t}&=
  -\frac{\hbar^2}{2m}\frac{\partial^2\Psi_2}{\partial x^2}
  +U(x)\Psi_2 + g_2|\Psi_2|^2\Psi_2 + g_{12}|\Psi_1|^2\Psi_2
  -\frac{\hbar\Omega}{2}\Psi_1 + \delta \Psi_2,
 \end{aligned}
\label{2GP}
\end{equation}
where $\Psi_1$ and $\Psi_2$ are macroscopic wave functions corresponding
to the first and the second hyperfine states, respectively,
$U(x)$ is the lattice potential, $g_1$ and $g_2$ are the nonlinearity parameters corresponding to 
intra-component interaction, $g_{12}$ is the nonlinearity parameter quantifying the inter-component
interaction, $\Omega$ is the Rabi frequency being the rate of the inter-component transitions, and $\delta$ is the transition detuning.
Hereafter we shall consider the case of $\delta=0$.

In the absence of Rabi coupling, $\Omega=0$, spatial dynamics is controlled 
by the lattice height and nonlinearity parameters.
The lattice height determines the rate of tunneling between neighbouring sites.
Strong nonlinearity can significantly reduce the wavepacket expansion due 
to spatial self-trapping \cite{TrSmerzi,Anker}.
In addition,
dynamics of two-component condensate mixtures 
with repulsive interactions (positive nonlinearity parameters)
between atoms  qualitatively depends on whether this mixture is miscible or not.
 The condition of immiscibility can be formulated as follows
\begin{equation}
g_{12}^2>g_1g_2,
 \label{immiscible}
\end{equation}
i.~e. inter-component repulsion should prevail over intra-component repulsion.
Presence of the Rabi coupling can drive the mixture from miscible to immiscible regime
and vice versa \cite{Oberthaler2,Malomed}.
In experiments, each of the nonlinearity parameters $g_1$, $g_2$ and $g_{12}$ can be tuned via the Feshbach resonance.
As long as different components correspond to the same atoms, one can assume $g_1=g_2\equiv g$.

If the lattice is sufficiently deep, one can simplify the problem by means of the tight-binding approximation \cite{Ruostekoski,Kolovsky-Sib,JPB}.
In this way, it is assumed that the majority of energy is preserved in the first energy band,
and one can expand the wavefunctions $\Psi_1$ and $\Psi_2$ over first-band
Wannier states corresponding to the decoupled problem, i.~e. $\Omega=0$. 
One has to take care that the tight-binding approximation 
becomes essentially questionable if the criterion (\ref{immiscible}) is satisfied.
In the immiscible regime, initially uniform mixture undergoes 
fragmentation due to spatial separation of different components inside individual lattice sites
\cite{Ronen,Shrestha}, that is, the wavepackets occupying individual lattice sites
acquire essentially non-symmetric form and cannot be represented as
superposition of the first-band Wannier states.
Taking this into account, we set $g_{12}=0$
inferring the most unfavorable conditions for demixing.
Thus, Eqs. (\ref{2GP}) reduce to the 
coupled discrete Gross-Pitaevskii equations
\begin{equation}
\begin{aligned}
 i\hbar\frac{da_n}{dt}&=-J(a_{n-1} + a_{n+1}) + g|a_n|^2a_n -\frac{\hbar\Omega}{2}b_n,\\
 i\hbar\frac{db_n}{dt}&=-J(b_{n-1} + b_{n+1}) + g|b_n|^2b_n -\frac{\hbar\Omega}{2}a_n,
 \end{aligned}
\label{TB}
\end{equation}
where $a_n$ and $b_n$ are complex-valued amplitudes of the condensate wave function at the lattice site $n$,
$J$ is the hopping rate describing tunneling between neighbouring sites.
To our knowledge, the first systematic study of such models was presented
in \cite{Eilbeck}, where some particular analytical solutions were found.
Similar models were considered in \cite{Mazzarella,Maksimov}.
Hereafter we shall use the scaling of variables which corresponds to $\hbar=1$.

\subsection{Uncoupled lattice sites}\label{Uncoupled}
\pst

Firstly, let's consider the case of uncoupled lattice sites, $J=0$.
It corresponds to large values of the lattice potential height, when tunneling can be neglected.
In this case total occupation of each site 
\begin{equation}
  \rho_n=|a_n|^2 + |b_n|^2
\label{rho_n}
\end{equation}
is time-independent.
Substituting 
\begin{equation}
 a=|a_n|e^{i\alpha_n},\qquad b=|b_n|e^{i\beta_n}
\end{equation}
into (\ref{TB}), and after some simple algebra, we obtain the system of equations
\begin{equation}
\begin{aligned}
  \dot z_n=\sqrt{1-z_n^2}\sin\phi_n,\\
  \dot\phi_n=\Lambda_n z_n+\frac{z_n}{\sqrt{1-z_n^2}}\cos\phi_n,
  \end{aligned}
\label{zsys}
\end{equation}
where the dot denotes differentiation with respect to the rescaled time $t'=\Omega t$,
\begin{equation}
 z_n = \frac{|a_n|^2-|b_n|^2}{\rho_n}
 \end{equation}
is the normalized population imbalance corresponding to the $n$-th lattice site, 
\begin{equation}
\phi_n = \arg a_nb_n^*\equiv\alpha_n-\beta_n,
\label{phi_n}
\end{equation}
$\Lambda_n=g\rho_n/(\hbar\Omega)$. Equations (\ref{zsys}) originate from the classical Hamiltonian
\begin{equation}
 H(z_n,\phi_n) = \frac{\Lambda_n z_n^2}{2} - \sqrt{1-z_n^2}\cos{\phi_n}.
 \label{Ham}
\end{equation}
This expression is similar to the Hamiltonian of the nonlinear pendulum. The main difference is the presence 
of the factor $(1-z_n^2)^{1/2}$. As this factor depends on $z_n$, the phase space structure 
corresponding to the Hamiltonian (\ref{Ham}) crucially depends on the parameter $\Lambda$.
If $\Lambda_n<1$, i.~e. in the case of weak nonlinearity, the phase portrait corresponding to Eqs.~(\ref{zsys}) contains 
a family of center fixed points located at $z_n=0$, $\phi_n=\pi m$, $m=1,2$.
Any trajectory of (\ref{zsys}) is rotation in phase space around one of the fixed points \cite{Oberthaler1}.
It corresponds to oscillations of the population imbalance $z_n$ with zero mean.
This dynamical regime can be referred to as the Rabi regime \cite{Leggett}.

If $\Lambda_n>1$, the fixed point located at $z_n=0$, $\phi_n=\pi$  undergoes the pitchfork
bifurcation, that is, it transforms into three fixed points, one saddle and two centers.
The saddle fixed point is located at $z_n=0$, $\phi_n=\pi$, i.~e. at the same place as 
the center fixed point before the bifurcation.
The ``new'' center fixed points are located at $z_n=\pm\sqrt{1-\Lambda_n^{-2}}$, $\phi_n=\pi$.
Each of them is surrounded by the domain of population imbalance
oscillations with nonzero mean, that is,
one component dominates over another.
This regime can be referred to as the {\it internal self-trapping}.
This phenomenon was earlier observed for the cases of interacting \cite{Oberthaler1} and non-interacting \cite{JETP09} atoms.
Further growth of the intercation strength $g$ moves the center fixed points towards limiting values
of the population imbalance, $z_n=\pm 1$, with increasing of the phase space area corresponding to the internal self-trapping.

\section{Spatial dynamics of the mixture in the optical lattice}\label{Spatial}
\pst

For moderate values of the optical potential height, $U_0\sim 10$ recoil energies or less, 
tunneling between lattice sites becomes meaningful and has to be taken into account.
 Tunneling results in spatial spreading of an atomic wavepacket along the lattice, thereby reducing 
local condensate density.
However, if the interaction energy 

\begin{equation}
 E_{\text{int}} = g\sum\limits_n (|a_n|^4 + |b_n|^4)
 \label{Einter}
\end{equation}
exceeds some threshold value, there can occur spatial self-trapping, 
when wavepacket depletion becomes very slow due to formation of steady soliton-like states \cite{Anker}.
Spatial spreading also depends on the wavepacket form, in particular,
on its width and spatial modulation \cite{TrSmerzi,Politi}.
All the computations presented in this paper were performed with $J=1$. 

In order to study the effect of spatial modulation, 
let's consider the initial condition of the following form:
\begin{equation}
 a_n(t=0) = A\exp\left[-\frac{n^2}{4\sigma^2}\right]\cos{\pi pn},\quad
 b_n=0,
\label{init}
\end{equation}
where $p\in [0:2]$, $-N\le n\le N$, $\sigma=10$,
the factor $A$ is determined by the normalization condition
\begin{displaymath}
 \sum\limits_{n=-N}^{N} |\psi_n|^2 = 1.
\end{displaymath}
Throughout this paper we perform computations for the lattice with 10001 sites, i.~e. $N=5000$.
With this normalization, information about number of condensed atoms
and s-wave scattering length
is hidden in the nonlinearity parameter $g$.
In order to distinguish different regimes of nonlinearity,
it is convenient to set a value of the interaction energy 
$E_{\text{int}}$, and then compute the corresponding value of the nonlinearity 
parameter by means of the formula
\begin{equation}
  g = \frac{E_{\text{int}}}{\sum\limits_{n=-N}^{N} |a_n(t=0)|^4}.
\end{equation}
In the absence of Rabi coupling, wavepacket dynamics qualitatively depends on the parameter $p$
which describes phase configuration of the initial state \cite{TrSmerzi,Politi}.
The case of $p=0$ corresponds to the same wavepacket phases for all lattice sites,
while the case of $p=1$ means the checkboard-like ordering of phases 0 and $\pi$.
The case of $p=0.5$ also means the checkboard sequence, but the sites with phases
0 and $\pi$ are separated by unoccupied sites.
 If the wavepacket width is sufficiently large, $\sigma\sim 10$,
 the strongest spreading is expected in the absence of spatial modulation of the initial state, 
 i.~e. for $p=0$ or $p=2$,
 while the strongest spatial self-trapping is anticipated for $p=0.5$ or $p=1.5$.
 The intermediate case of $p=1$ is the most favorable from the viewpoint
of the breather formation \cite{TrSmerzi}.

\begin{figure}[!htb]
\centerline{
\includegraphics[width=0.4\textwidth]{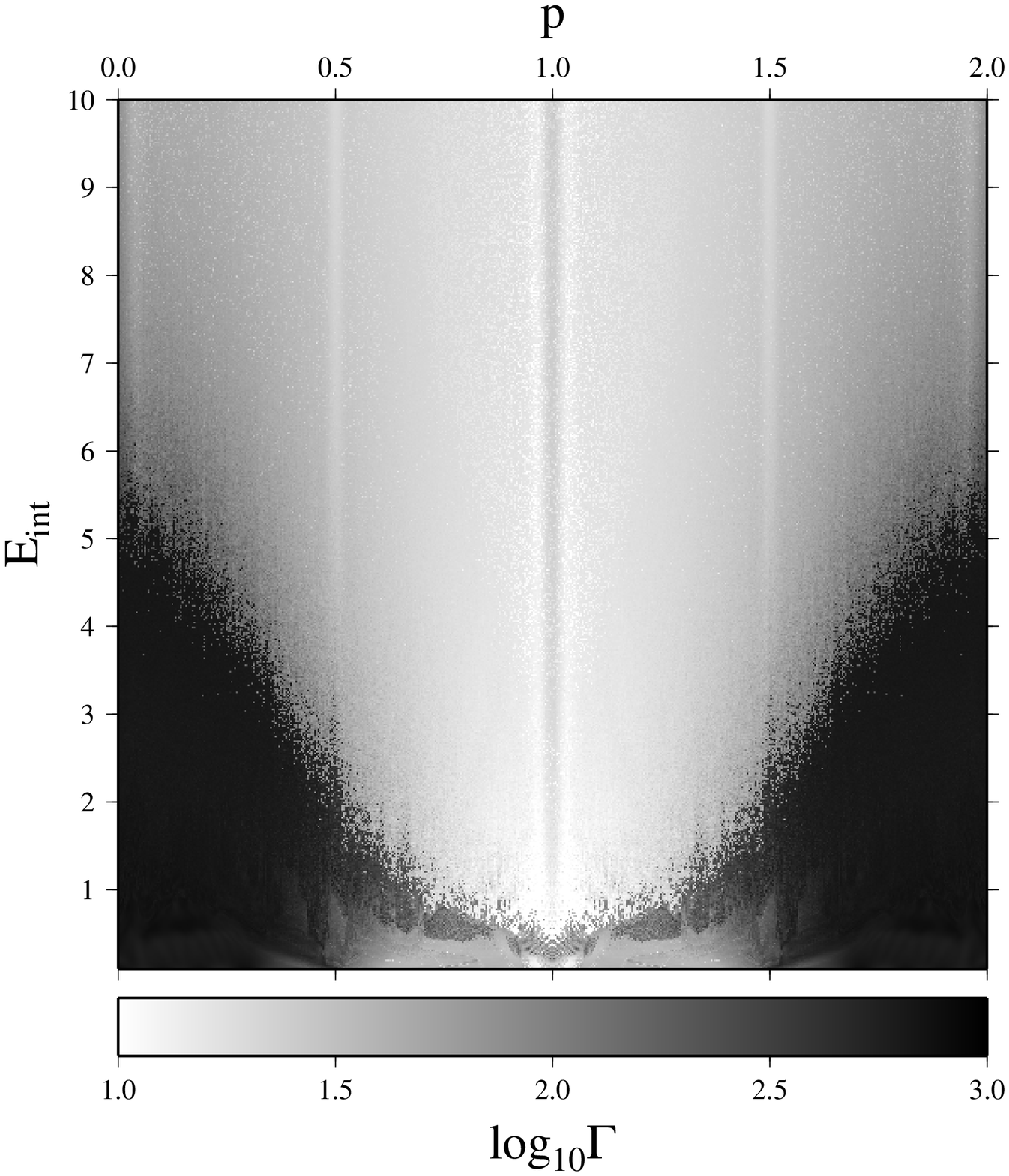}
\includegraphics[width=0.4\textwidth]{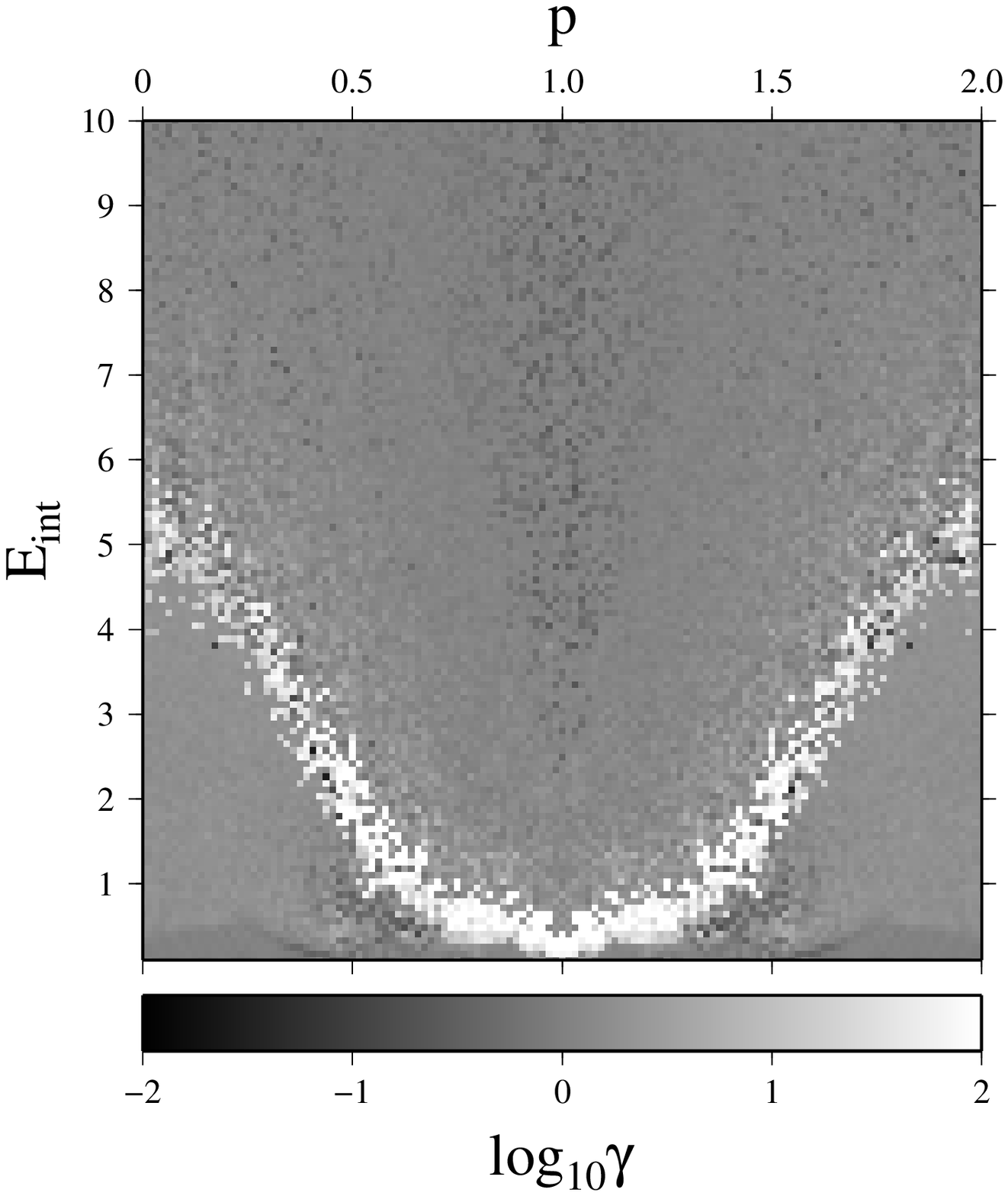}}
\caption{Left: the map of participation ratio values computed at $t=1000\pi$. 
Right: the map of relative differences between the paticipation ratios computed with
and without Rabi coupling. Color scales for both plots are presented below.} 
\label{fig-maps}
\end{figure}

BEC expansion along the lattice can be quantified by 
position variance
\begin{equation}
 \sigma = \sqrt{\sum\limits_n n^2\rho_n - \left(\sum\limits_n n\rho_n \right)^2},
\end{equation}
or by the participation ratio
\begin{equation}
 \Gamma=\frac{1}{\sum\limits_{n} \rho_n^2},
\end{equation}
where $\rho_n$ is given by (\ref{rho_n}).
The participation ratio is approximately equal to the number 
of lattice sites which are efficiently occupied.

Left panel of figure \ref{fig-maps} represents the map of participation ratio values computed
at $t=1000\pi$. 
Coordinates of the map are the parameter $p$ and the interaction energy $E_{\text{int}}$.
There is sharp contrast between ``dark'' regions corresponding to fast condensate depletion due to spatial diffusion,
and ``light'' region where the condensate concentration remains sufficiently high.
The light region becomes broader with increasing $E_{\text{int}}$, reflecting onset of spatial self-trapping.
It involves notable vertical lines corresponding to $p=0.5,1.0$ and $1.5$. The lines at $p=0.5$ and $1.5$ 
correspond to smaller spreading as compared with the nearest background, while the line at $p=0$ looks a little more
dark and corresponds to larger spreading. These results are qualitatively consistent with theoretical predictions
of \cite{TrSmerzi}. To examine the difference brought in by the Rabi coupling,
it is reasonable to compare the above results with the results obtained without the Rabi coupling.
It is made in the right panel of Fig.~\ref{fig-maps},
where we plot the map of the the quantity defined as
%
\begin{equation}
 \gamma = \frac{\Gamma(\Omega=1,t=1000\pi)}{\Gamma(\Omega=0,t=1000\pi)}.
 \label{gamma}
\end{equation}
If the Rabi coupling results in increasing (decreasing) of condensate spreading,
then this quantity is larger (smaller) than 1.
The most prominent difference is observed near the boundary 
separating ``dark'' and ``light'' parameter regions 
in the left panel. Near this boundary, the Rabi coupling 
slightly extends the parameter area corresponding to fast spatial diffusion.
This can be understood as reduction of intra-component interactions
due to component interconversion. Indeed, the interaction energy oscillates in the 
presence of the Rabi coupling, and its time-averaged value becomes smaller.
On the other hand, one can see a cloud of dark points 
in the region corresponding to $p=0$ and large values of $E_{\text{int}}$.
It means that, in the regime of strong nonlinearity, Rabi coupling 
can weaken the spreading of a wavepacket.


%
\begin{figure}[!htb]
\centerline{\includegraphics[width=0.48\textwidth]{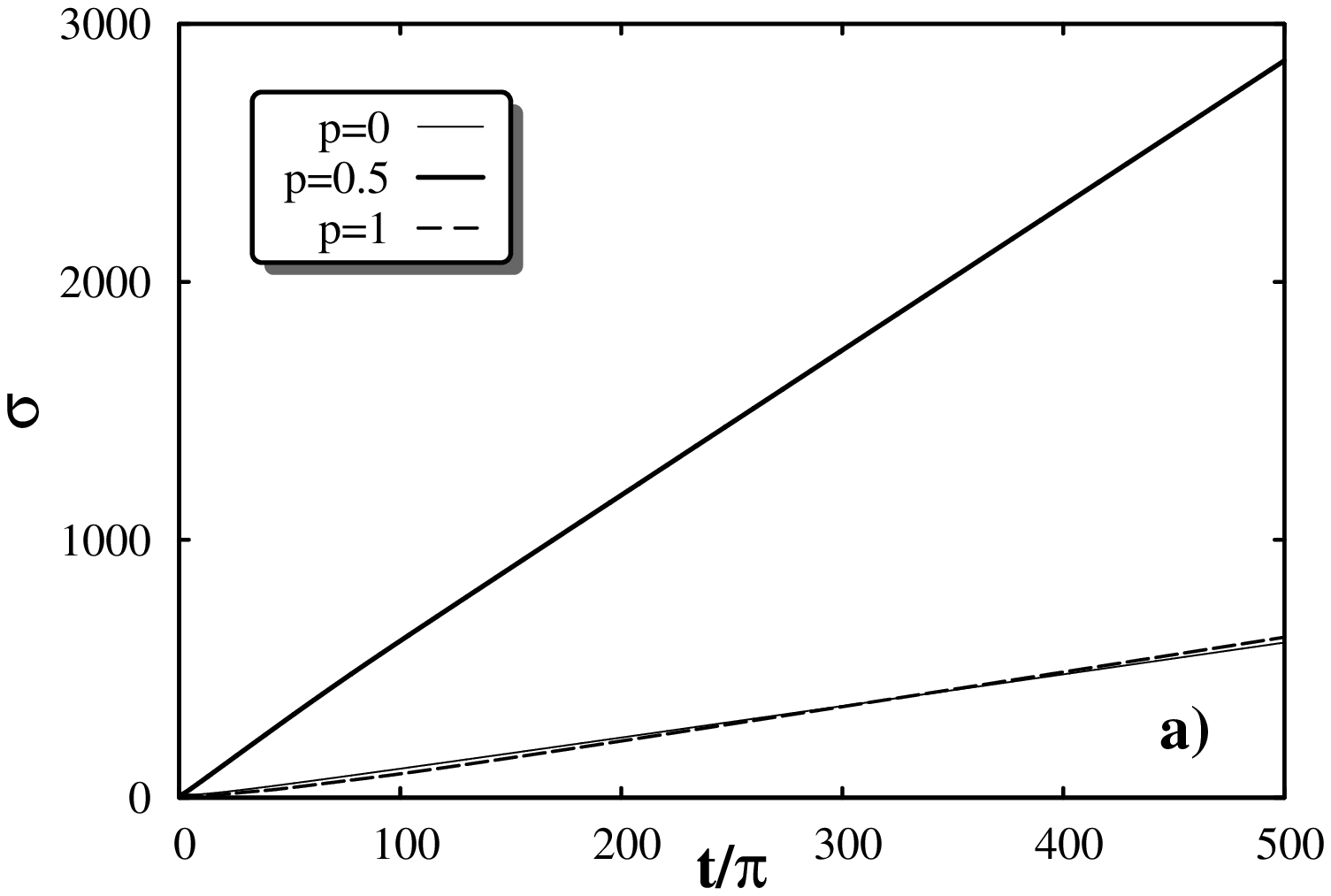}
\includegraphics[width=0.48\textwidth]{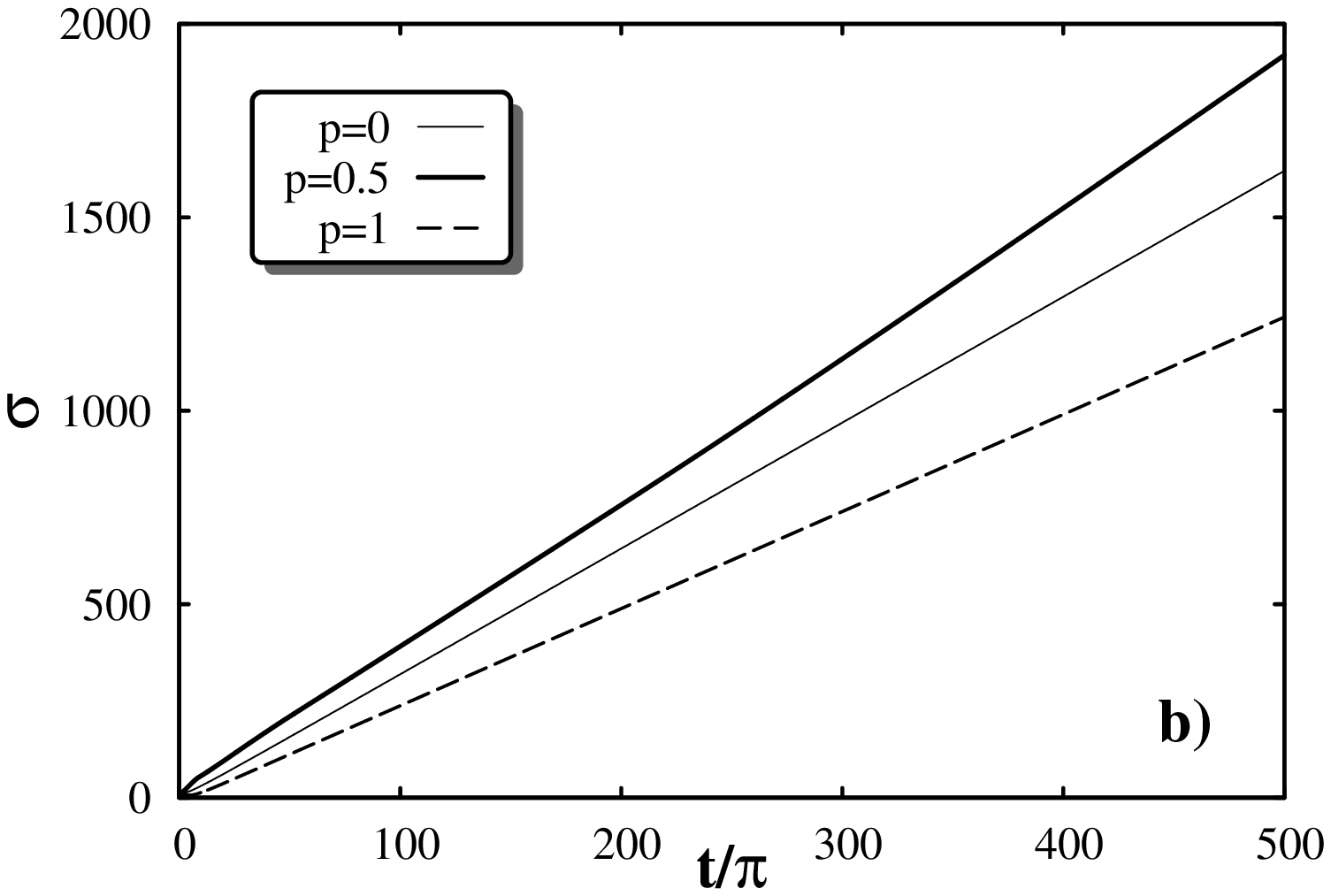}
}
\caption{Position variance $\sigma$ as function of time for different values of the interaction energy:
a) $E_{\text{int}}=0.1$, b) $E_{\text{int}}=1$. 
Rabi frequency $\Omega=1$ for all cases.
}
\label{fig-sigma}
\end{figure}

\begin{figure}[!htb]
\centerline{\includegraphics[width=0.48\textwidth]{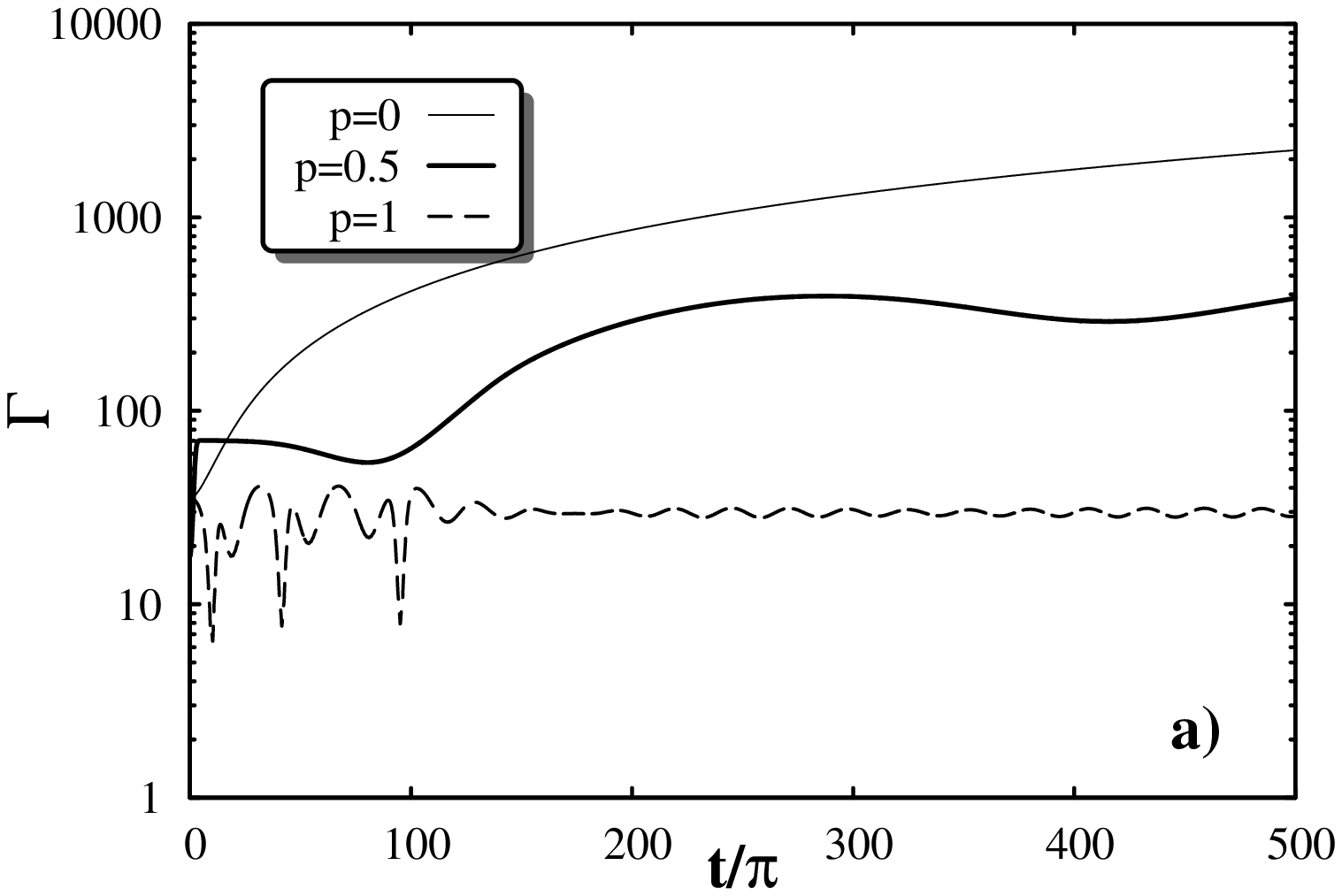}
\includegraphics[width=0.48\textwidth]{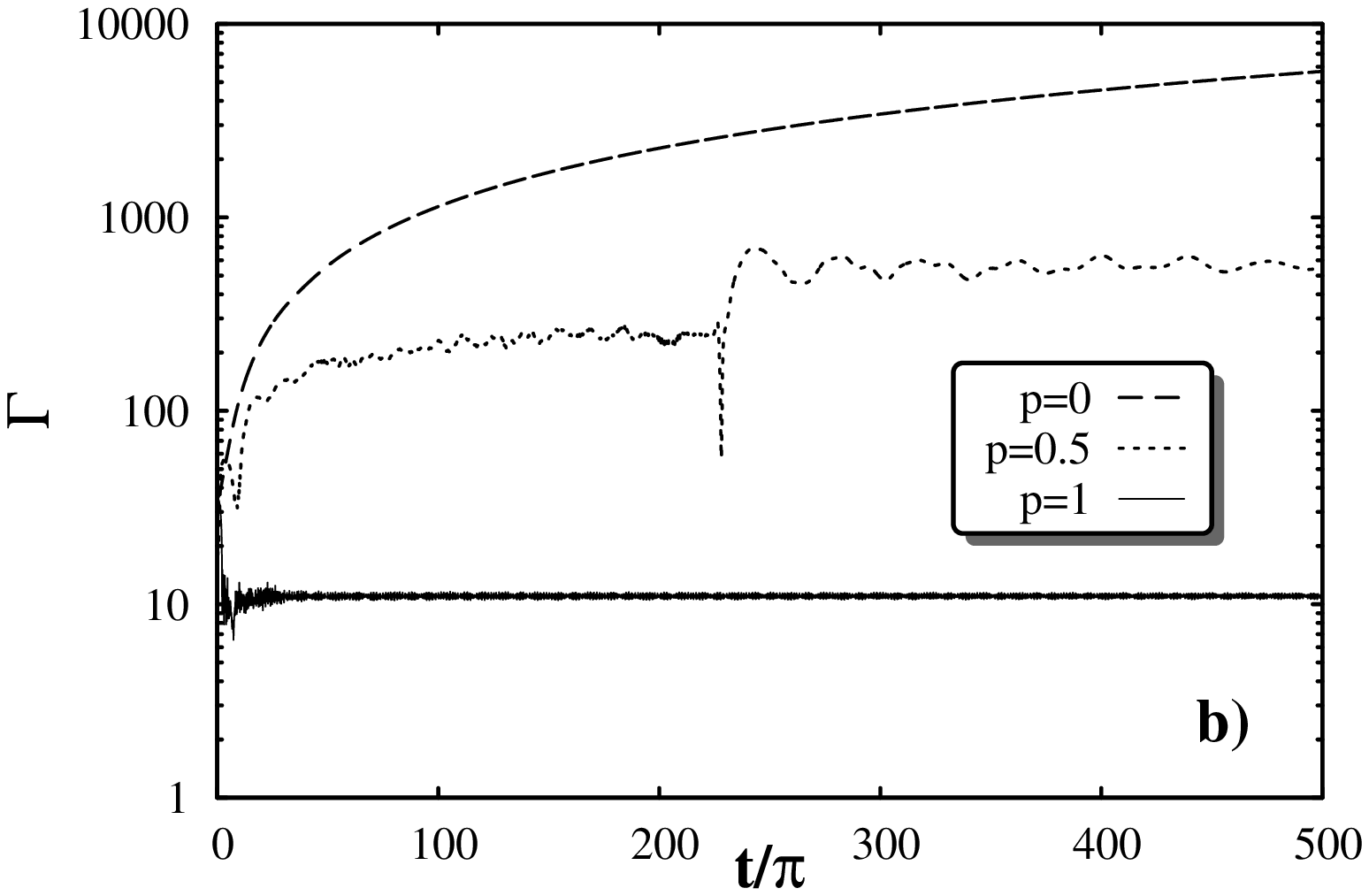}
}
\caption{Participation ratio as function of time for different values of the interaction energy:
a) $E_{\text{int}}=0.1$, b) $E_{\text{int}}=1$. 
Rabi frequency $\Omega=1$ for all cases.
} 
\label{fig-pr}
\end{figure}

Figures \ref{fig-sigma} and \ref{fig-pr} illustrate the process of wavepacket 
spreading for different values of the interaction energy.
In particular, we shall consider the cases of $E_{\text{int}}=0.1$ and
 $E_{\text{int}}=1$, referring to them as the regime of weak and
moderate intra-component interactions, respectively.
We consider initial wavepackets with $p=0$, 0.5 and 1.
Let's begin with considering Fig.~\ref{fig-sigma} demonstrating
time dependence of the position variance.
In the regime of weak intra-component interactions, $E_{\text{int}}=0.1$,
the variance grows on the ballistic manner (see Fig.~\ref{fig-sigma}a).
Notably, rate of the growth for the case of $p=0.5$ is remarkably larger than for $p=0$ and $p=1$.
Ballistic expansion is also observed in the case of moderate intra-component interactions, $E_{\text{int}}=1$,
but the difference between different values of $p$ becomes not so apparent (see Fig.~\ref{fig-sigma}b).

Evolution of participation ratio is presented in Fig.~\ref{fig-pr}. 
One can see that number of sites occupied
by the wavepacket with $p=1$ 
 varies weakly for both regimes of nonlinearity.
This can be thought of as a signature of self-trapping and
breather formation.
The opposite situation is observed in the case $p=0$, when
to the most fastest growth of the participation ratio
anticipates the strongest tendency to diffusion.
In the case of $p=0.5$, fast growth of $\sigma$ in the regimes
of weak or moderate intra-component interactions is accompanyied by 
relatively weak growth of the participation number.
This indicates on the formation of ballistically moving breathers.

\section{Internal dynamics}\label{Internal}

\subsection{Weak intra-component interaction}\label{Weak}
\begin{figure}[!htb]
\centerline{\includegraphics[width=0.49\textwidth]{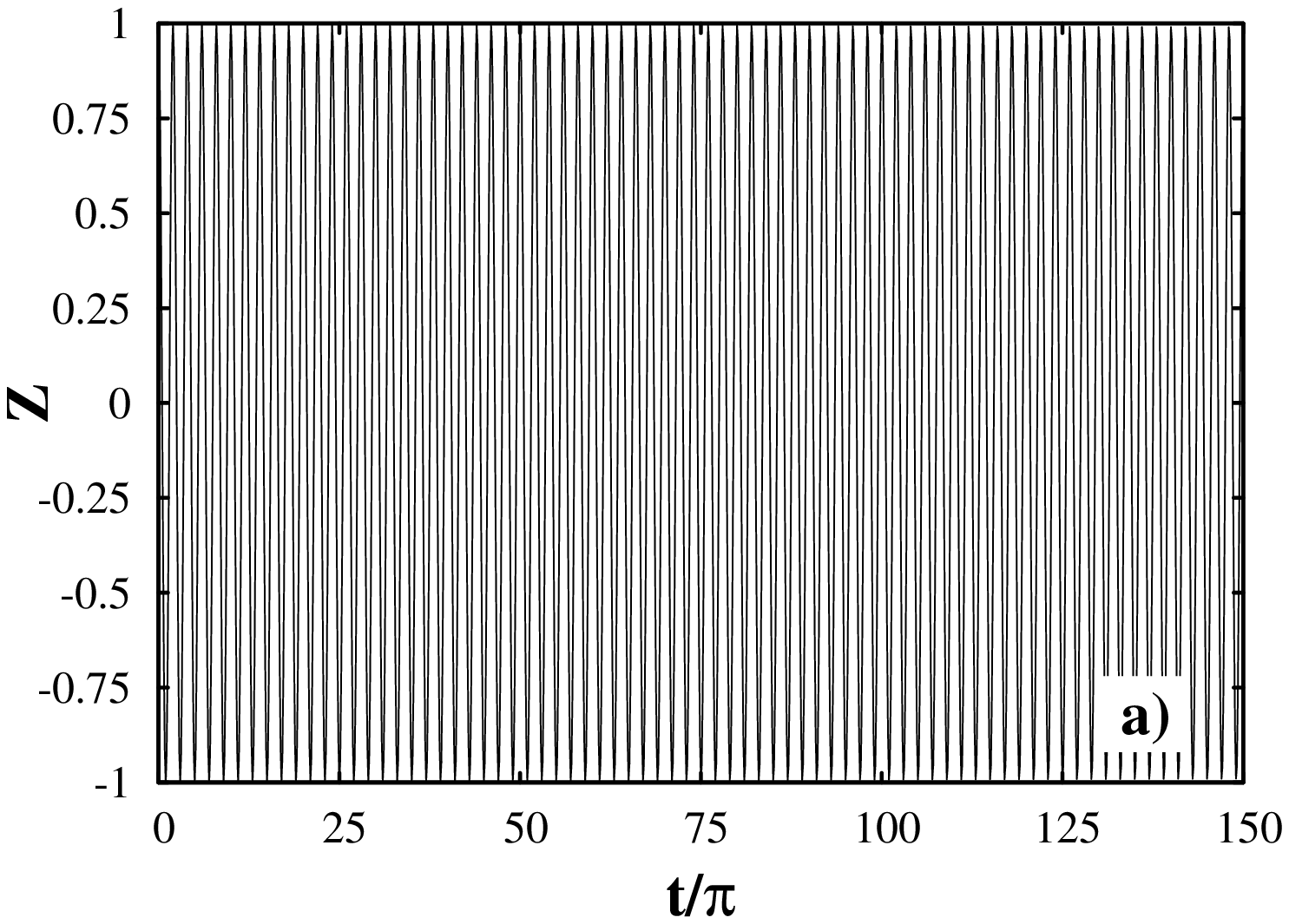}
\includegraphics[width=0.49\textwidth]{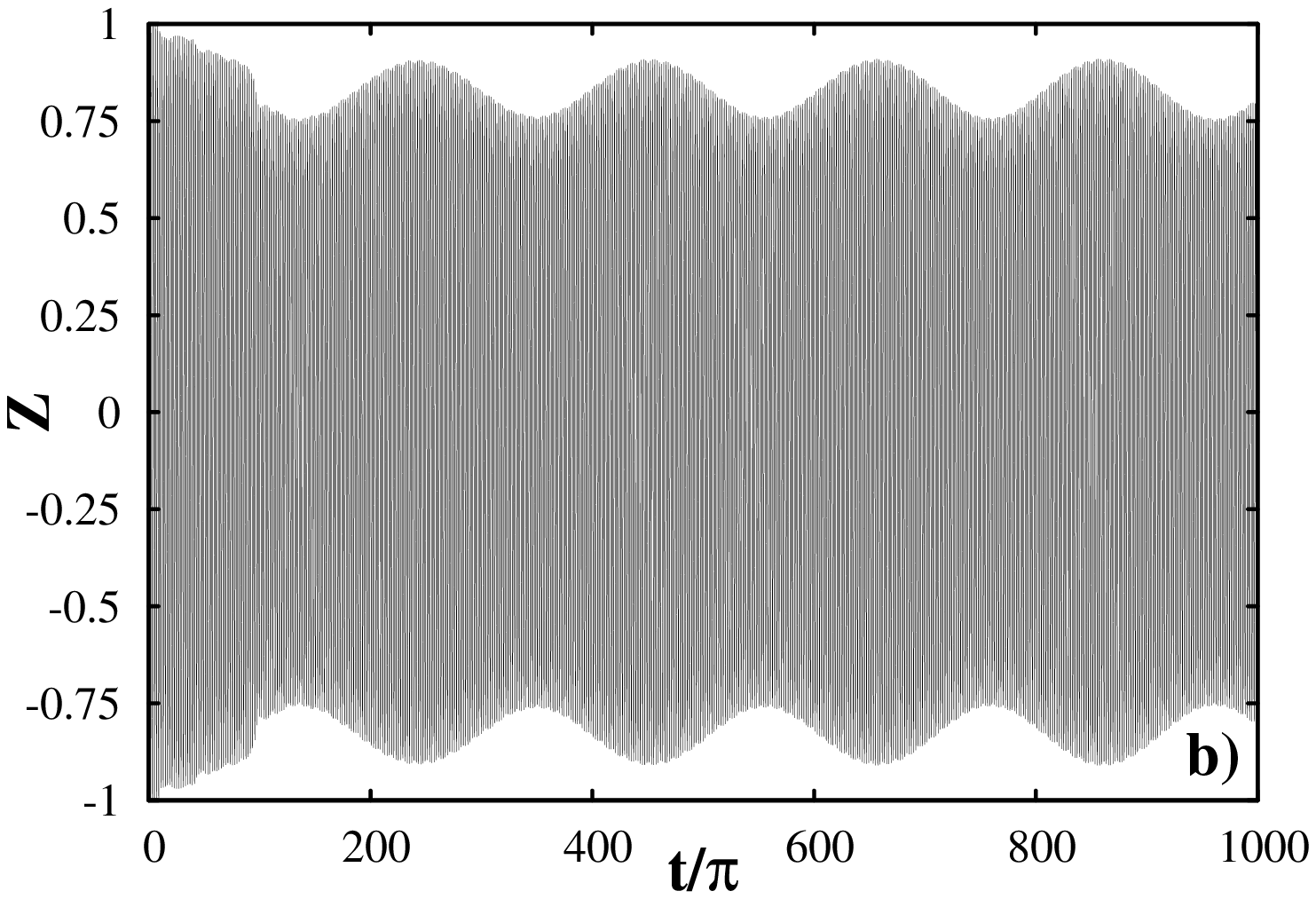}
}
\caption{
Total population imbalance as function of time for the case of weak intra-component interactios,
$E_{\text{int}}=0.1$. a) $p=0$, b) $p=1$.
}
\label{fig-z-weak}
\end{figure}

Let's consider how spatial dynamics of the mixture influences inter-level transitions.
For this purpose, it is reasonable to examine evolution of the total population imbalance (TPI)
\begin{equation}
Z=\sum\limits_n \rho_nz_n = \sum\limits_n |a_n|^2 - \sum\limits_n |b_n|^2.
 \label{Z}
\end{equation}
Firstly, let's consider the case of weak intra-component interaction. 
Figure \ref{fig-z-weak} illustrates dependence of TPI on time for $p=0$ and $p=1$.
In both these cases, TPI undergoes oscillations which are typical
for the Rabi regime. 
In the case of $p=0$,
these oscillations are almost completely 
coherent and their amplitude is close to the maximal value. 
It means that linear Rabi coupling provides complete and synchronous component interconversion.
The same behavior is observed in the case of the initial state with $p=0.5$.
In the case of $p=1$ the situation is qualitatively different, and the amplitude of TPI oscillations
varies periodically with time, as is demonstrated
in Fig.~\ref{fig-z-weak}b. This difference comes from the difference in spatial behavior.
In the case of $p=0$, spatial wavepacket expansion is accompanied by fast growth of the participation ratio.
Therefore, local condensate density rapidly decreases and
intra-component interaction becomes negligible.
This results in decoupling of translational and internal degrees of freedom, that is,
all the lattice sites undergo synchronous Rabi oscillations.
In the case of $p=1$, the condensate depletion is much weaker (see Fig.~\ref{fig-pr}a)
due to formation of two distinct pairs of breathers. 
Each pair consists of identical breathers disposed 
symmetrically with respect to the center of the initial state. i.~e., $n=0$.
Difference in condensate densities of the pairs results
in the difference in frequencies of Rabi oscillations, therefore,
TPI varies quasiperiodically like sum of two harmonic functions.

\begin{figure}[!htb]
\centerline{\includegraphics[width=0.33\textwidth]{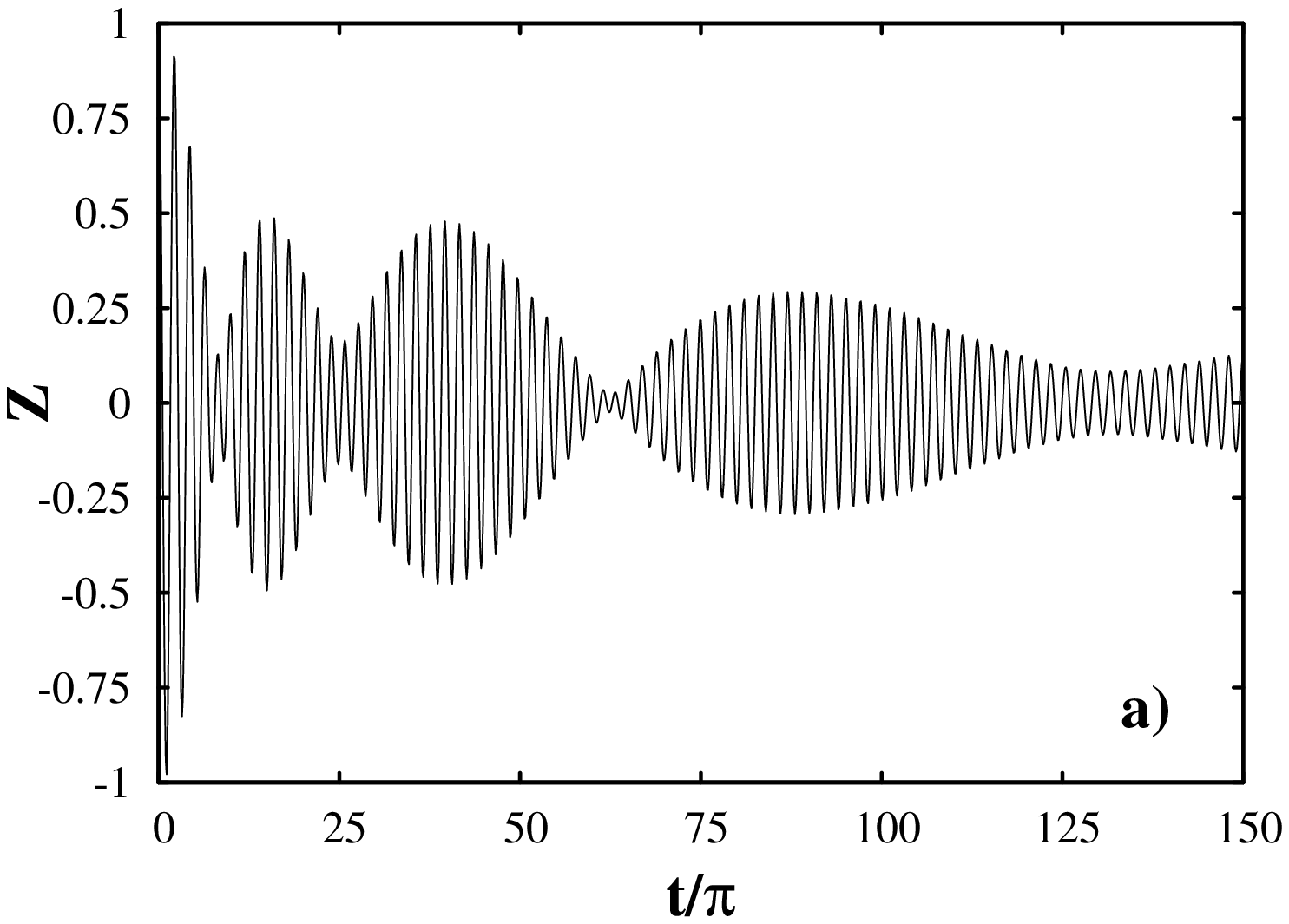}
\includegraphics[width=0.33\textwidth]{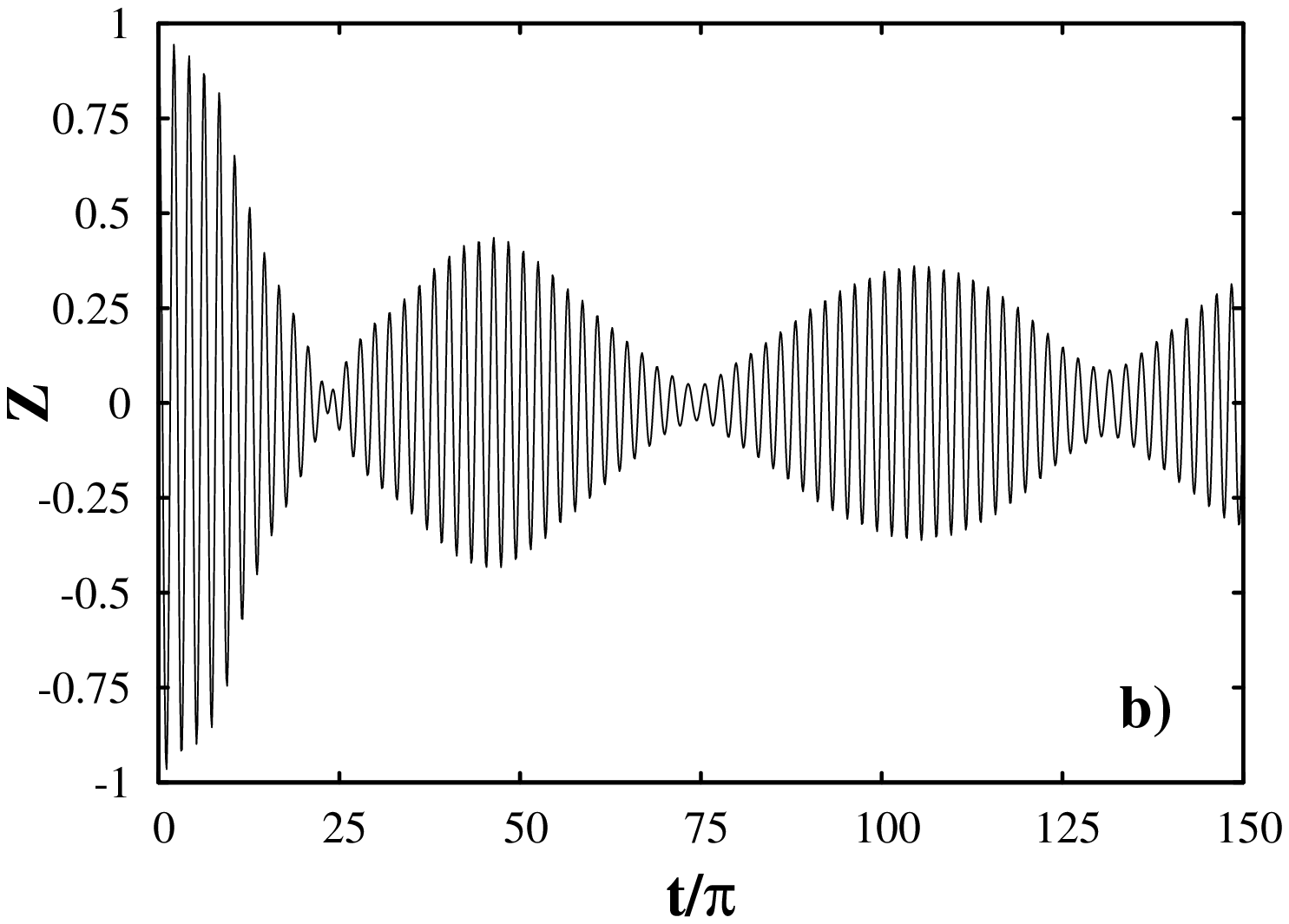}
\includegraphics[width=0.33\textwidth]{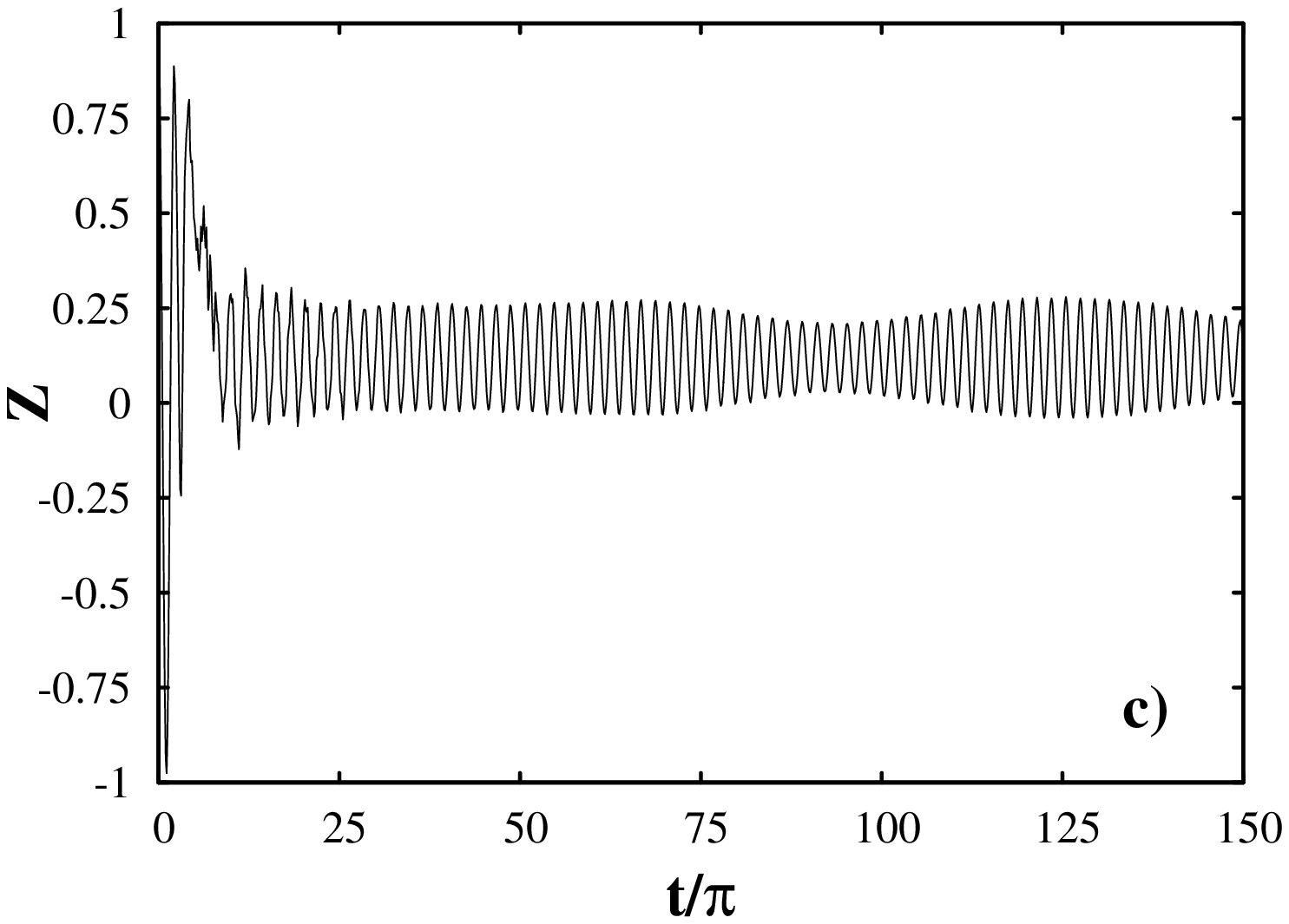}
}
\caption{
Population imbalance as function of time for the case of moderate intra-component interactions,
$E_{\text{int}}=1$.
a) $p=0$, b) $p=0.5$, c) $p=1$.
} 
\label{fig-z-mod}
\end{figure}

\begin{figure}[!htb]
\centerline{\includegraphics[width=0.6\textwidth]{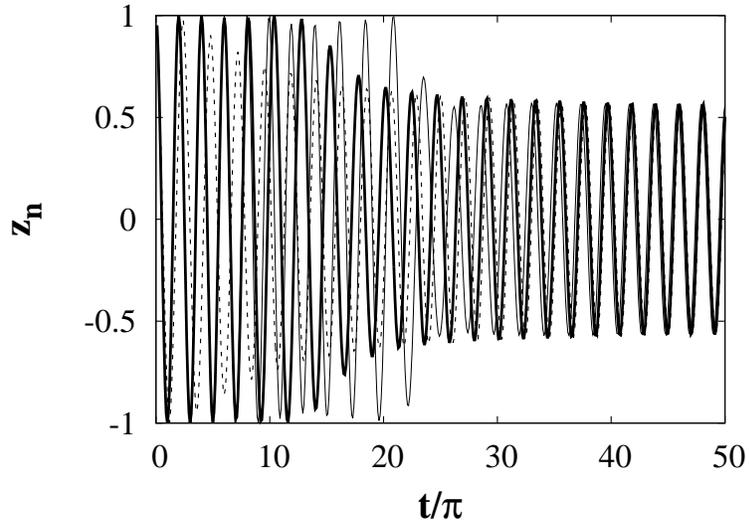}
}
\caption{Synchronization of
Rabi oscillations corresponding to 
the sites $n=0$ (dashes), $n=30$ (thick solid) and
$n=50$ (thin solid).
} 
\label{fig-sync2}
\end{figure}

\begin{figure}[!htb]
\centerline{
\includegraphics[width=0.5\textwidth]{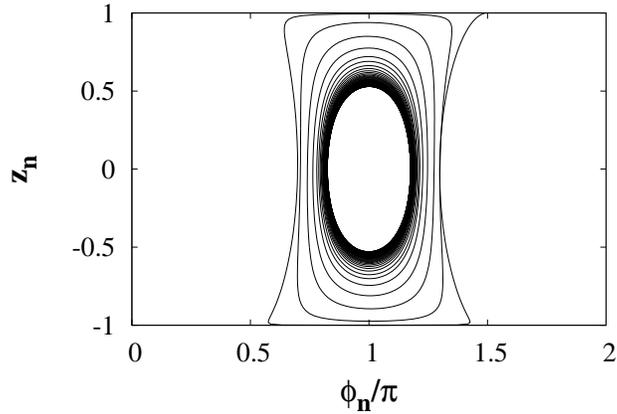}}
\caption{Typical phase portrait describing inter-component transitions inside an individual lattice site for $p=0$.
$\phi_n$ is given by (\ref{phi_n}).
} 
\label{fig-cell}
\end{figure}

\begin{figure}[!htb]
\centerline{\includegraphics[width=0.6\textwidth]{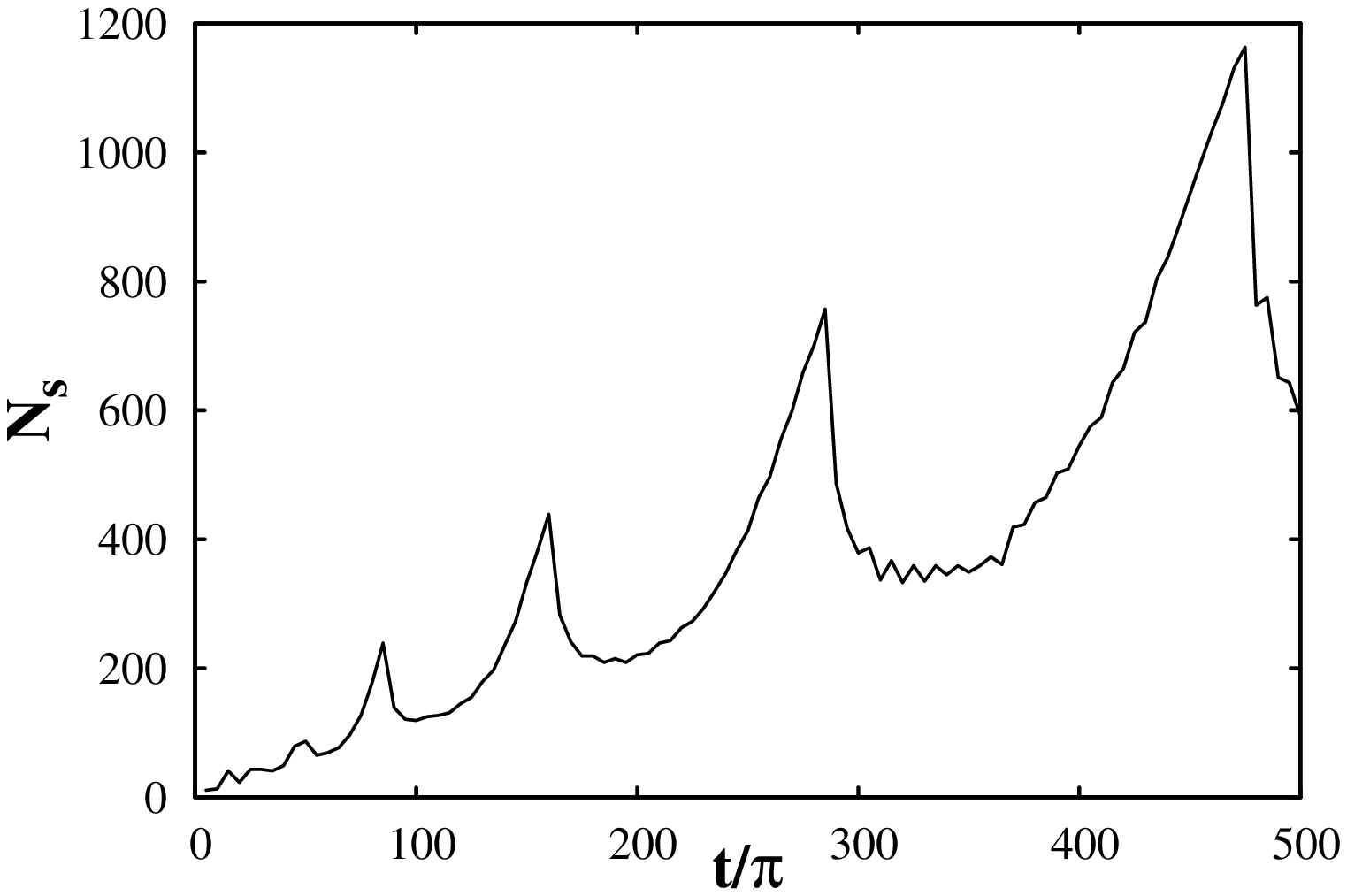}
}
\caption{
Number of synchronized sites as function of time.
} 
\label{fig-synclength}
\end{figure}

\subsection{Moderate intra-component interaction}\label{Moderate}

Increasing of nonlinearity in (\ref{TB}) 
gives rise to anharmonicity in inter-level transitions.
Impact of nonlinearity depends on local density of the condensate,
therefore, frequencies of Rabi oscillations at different lattice sites don't coincide.
The frequency offsets destroy the synchronization of Rabi oscillations.
As a consequence, the resulting TPI oscillations become the superposition 
of many independent oscillations running inside individual lattice sites, and the maximal amplitude
of TPI oscillations is significanltly lower than in the case of weak intra-component intercation.

However, if the intra-component interaction is not very strong,
large fraction of the condensate near the center of the initial state
can maintain synchronization, giving rise to a large domain with coherent Rabi dynamics.
This regime is realized in the case of the initial state with $p=0$ (see Fig.~\ref{fig-z-mod}a).
As it follows from Fig.~\ref{fig-sync2}, 
there occurs spontaneous synchronization of Rabi oscillations corresponding to different lattice sites.
Phase portraits in the $z_n$--$\phi_n$ plane,
constructed for individual lattice sites,
are very similar to each over and look as is demonstrated in Fig.~\ref{fig-cell}.
The marked common feature is the presence of the limit cycle whose phase space location
is nearly the same for all synchronized sites. This may appear somewhat surprising.  Indeed,
limit cycles can occur only in the presence of dissipation while the equations 
(\ref{TB}) don't contain dissipative terms. 
However, the role of dissipation can be played by one-way energy transfer between degrees of freedom.
In our case, energy from the internal degree of freedom, associated with the Rabi coupling,
is absorbed by the external degrees of freedom, namely spatial wavepacket motion and 
intra-component interaction. Hence, there arises synchronization inherent of non-Hamiltonian systems.

Taking into account that the domain of synchronized Rabi oscillations is placed in the center of the lattice, 
one can define the criterion of synchronization as inequality
\begin{equation}
|z_n(t) - z_0(t)|\le C,\quad -n'\le n\le n'.
\label{sync_crit} 
\end{equation}
Here $C$ is some small constant. 
We set $C=0.1$ in order 
to estimate number of synchronized sites
\begin{equation}
 N_{\text{s}}=2n'+1,
\end{equation}
Time dependence of $N_\text{s}$ is presented in Fig.~\ref{fig-synclength}.
It is demonstrated that $N_\text{s}$ undergoes pulsations with growing
amplitude. It implies that some lattice sites 
can exit the synchronized regime and then
return into it. This phenomenon indicates on the presence
of wave-like excitations inside the domain of synchronization, reminiscent of magnons.
Comparison with Fig.~\ref{fig-pr}b allows one to link the growth of pulsations amplitude
with increasing of participation ratio, that is, 
 more and more lattice sites become synchronized
 as the condensate expands along the lattice.

In the case of the initial state with $p=0.5$, TPI also undergoes
oscillations with decaying amplitude, 
as is shown in Fig.~\ref{fig-z-mod}b. However, there is no synchronization, and the amplitude decay 
is associated with dephasing of Rabi oscillations running at different sites. 

\begin{figure}[!htb]
\centerline{
\includegraphics[width=0.4\textwidth]{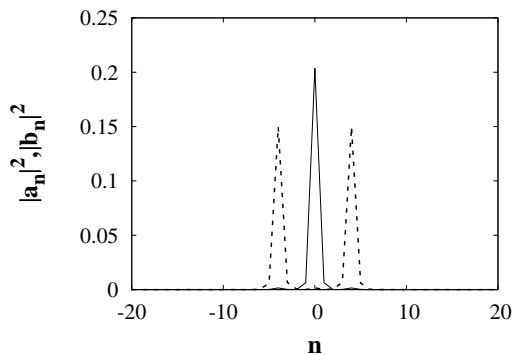}}
\caption{
Squared modulo of wavefunctions corresponding to the first (solid) and second (dotted) components at $t=500\pi$.
}
\label{fig-field-mod}
\end{figure}

Initial state with $p=1$ behaves in a qualitatively different way as compared with $p=0$ and $p=0.5$.
As one can see in Fig.~\ref{fig-z-mod}c, TPI undergoes oscillations with non-zero mean,
i.~e. the first component of the mixture prevails over the second one.
It indicates on the onset of  internal self-trapping, the effect discussed in Section~\ref{Uncoupled}.
In the absence of coupling between sites, internal self-trapping is linked to the bifurcation
of fixed points according to the scenario described in Section \ref{Uncoupled}. 
However, one should expect that tunneling between neighbouring sites should destroy the phase coherence needed 
for the onset regular oscillations around displaced fixed points.
This means that the onset of the internal self-trapping should occur when tunneling is weak.
In this way, it is worth reminding that the initial state with $p=1$
is characterized by the strongest spatial self-trapping, that is,
the tunneling is suppressed due to formation of localized states like breathers or solitons.
These localized states are presented in Fig.~\ref{fig-field-mod}.
It should be noted that they are well separated in space, that is,
they don't affect each other. 
This means that inter-component dynamics of every such state can be fairly described
within a single-site approximation corresponding to the Eqs.~(\ref{zsys}).
\begin{figure}[!htb]
\centerline{
\includegraphics[width=0.48\textwidth]{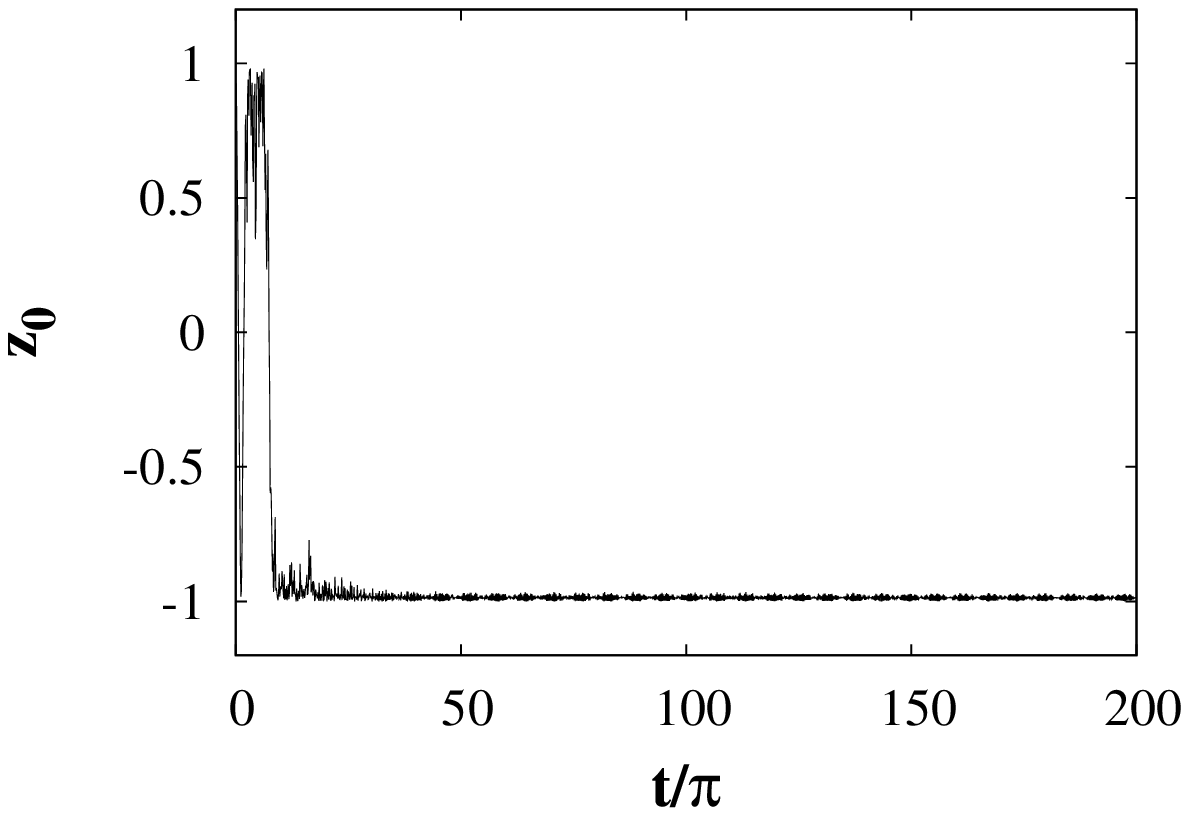}
\includegraphics[width=0.48\textwidth]{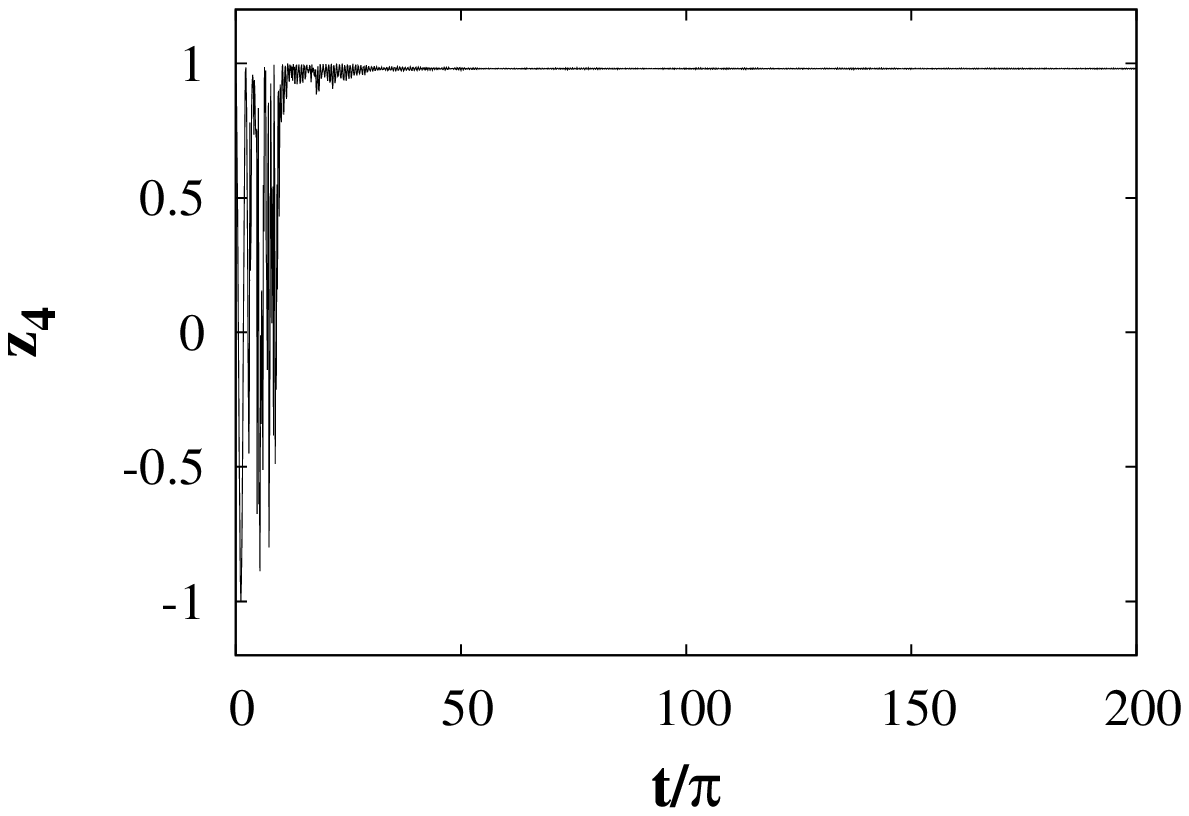}
}
\caption{
Population imbalances corresponding to the lattice sites $n=0$ (left figure)
and $n=4$ (right).
}
\label{fig-zcells}
\end{figure}
Figure \ref{fig-zcells} illustrates population imbalance oscillations corresponding
to the center (left figure) and right (right figure) localized states depicted in Fig.~\ref{fig-field-mod}.
After short irregular transient regime, population imbalance becomes frozen in both cases, indicating 
the total dominance of one component. It means that the localized states
can be thought of as immiscible solitons.
Thus, it turns out that stable internal self-trapping occurs in the presence
of spatial self-trapping. 

\section{Summary}
\label{Summary}

The present work is devoted to dynamics of two-component BEC with linear inter-component coupling loaded
into the optical lattice. 
The main goal is to describe the interplay between spatial and internal degrees of freedom.
It is found that spatial mixture dynamics is readily controlled 
by the interaction strength and phase configuration of the initial state, i.~e. likewise in the case 
of one-component condensate \cite{TrSmerzi}.
If condensate undergoes fragmentation with occurrence of localized and isolated patterns like solitons or breathers, then the total population
imbalance is superposition of incoherent Rabi oscillations with different amplitudes and frequencies.
As the intra-component interaction energy exceeds
 some threshold, then Rabi oscillations corresponding to distinct localized patterns cease, that is, these patterns
 become immiscible solitons comprising only one component of the mixture.
In this case one observes simultaneously spatial and internal self-trapping.
It should be mentioned that formation of immiscible solitons is preceded by a short chaotic transient
and can be regarded as some kind of self-organization. 

In the case of the initial state without spatial modulation, Rabi oscillations corresponding to different lattice sites 
can be synchronized. If the intra-component interaction is weak, the synchronization is total and the condensate undergoes
Rabi oscillations as a whole. In the case of moderate intra-component interactions,   
Rabi oscillations at different sites have different frequencies, therefore total synchronization is destroyed.
However, there can arise spontaneous synchronization inside a large lattice domain. 
It is shown that synchronized oscillations correspond to the same limit cycle in phase space portraits corresponding
to individual lattice sites. Number of synchronized states strongly oscillates with time due the presence of 
wave-like excitations inside the domain of synchronization. On average, this number grows with time due to condensate expansion along the lattice.

Detailed study of the particular phenomena described in this paper, like spontaneous synchronization or formation
of immiscible solitons, will be the object of our future work.
In addition, we intend to go beyond the approximations used in derivation of Eqs.~(\ref{TB}) and verify
the results obtained with more realistic approaches. Firstly, one should use 
continuous mean-field approximation corresponding to the coupled Gross-Pitaevskii equations (\ref{2GP}).
This allows one to estimate the role of the inter-band Landau-Zener tunneling
which is ignored in the tight-binding approximation.
Another important advantage of the continuous mean-field approximation is the ability
to avoid limitations on the nonlinearity coefficients $g_{1}$, $g_2$ and $g_{12}$. These limitations are
required for validity of the tight-binding approximation and discussed
in Section \ref{Basic}.




\section*{Acknowledgments}
\pst
The authors acknowledge the financial support provided within
the RFBR project 12-02-31416, and the joint project of the Siberian and Far-Eastern
Branches of the Russian Academy of Sciences (project 12-II-SO-07-022).

\end{document}